\documentclass{article}
\usepackage{spconf,amsmath,graphicx}
\usepackage{enumitem}

\title{Revisiting Copy-move Forgery Detection by Considering Realistic Image with Similar but Genuine Objects}
\name{Ye Zhu$^{1,2}$ \qquad Tian-Tsong Ng$^{2}$ \qquad  Xuanjing Shen$^{1}$ \qquad  Bihan Wen$^{3}$}
\address{$^{1}$ College of Computer Science and Technology, Jilin University, Changchun, China.\\
$^{2}$ Situational Awareness Analytics, Institute for Infocomm Research, Singapore.\\
$^{3}$ Electrical and Computer Engineering and CSL, University of Illinois at Urbana-Champaign, IL, USA.}

\begin{document}
%
\maketitle
\begin{abstract}
Many images, of natural or man-made scenes often contain \emph{Similar but Genuine Objects} (SGO). This poses a challenge to existing \emph{Copy-Move Forgery Detection} (CMFD) methods which match the key points / blocks, solely based on the pair similarity in the scene. To address such issue, we propose a novel CMFD method using \emph{Scaled Harris Feature Descriptors} (SHFD) that preform consistently well on forged images with SGO. It involves the following main steps: (i) Pyramid scale space and orientation assignment are used to keep scaling and rotation invariance; (ii) Combined features are applied for precise texture description; (iii) Similar features of two points are matched and RANSAC is used to remove the false matches. The experimental results indicate that the proposed algorithm is effective in detecting SGO and copy-move forgery, which compares favorably to existing methods. Our method exhibits high robustness even when an image is operated by geometric transformation and post-processing.
\end{abstract}

\begin{figure*} [htb]
\centering
\includegraphics[width=3.5cm,height=3cm]{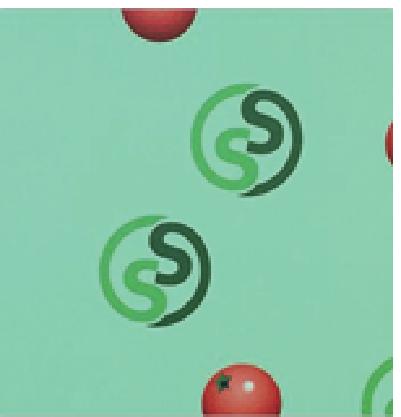}\hspace{0.02cm}
\includegraphics[width=3.5cm,height=3cm]{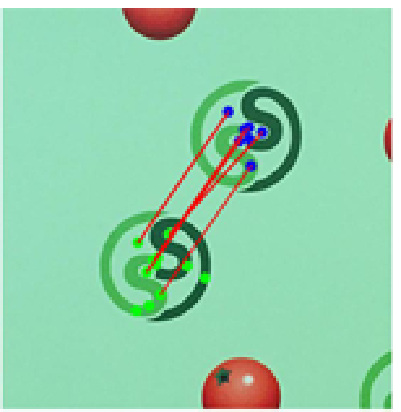}\hspace{0.02cm}
\includegraphics[width=3.5cm,height=3cm]{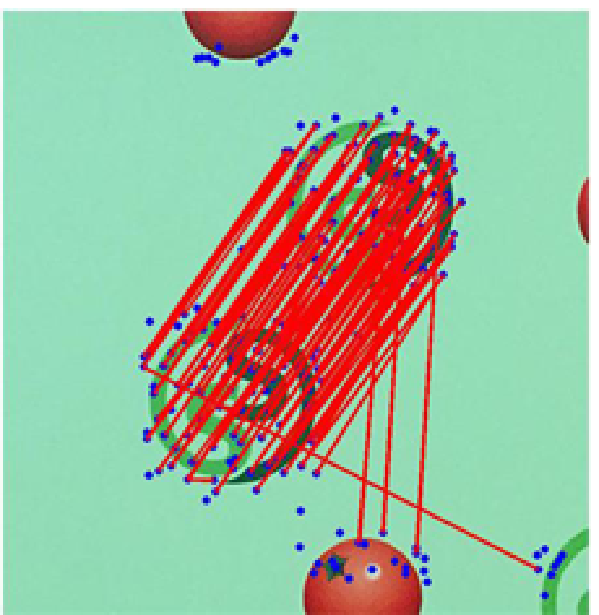}\hspace{0.02cm}
\includegraphics[width=3.5cm,height=3cm]{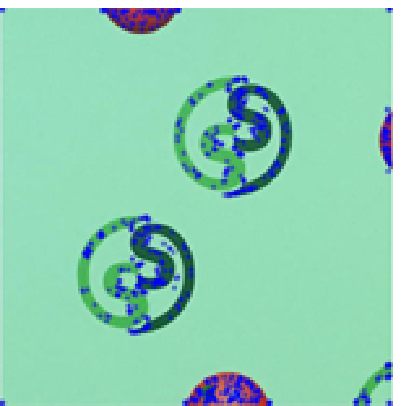}\hspace{0.02cm}

\vspace{.2cm}
\includegraphics[width=3.5cm,height=3cm]{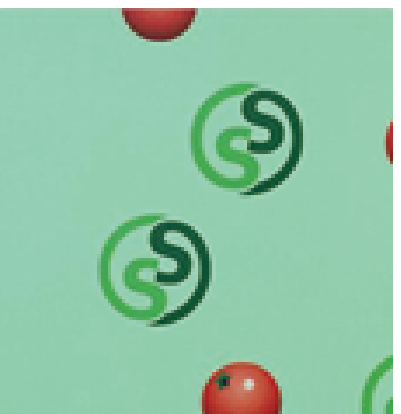}\hspace{0.02cm}
\includegraphics[width=3.5cm,height=3cm]{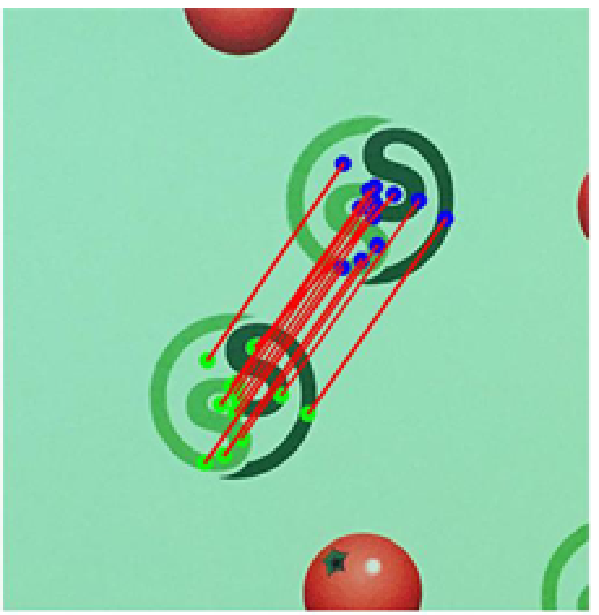}\hspace{0.02cm}
\includegraphics[width=3.5cm,height=3cm]{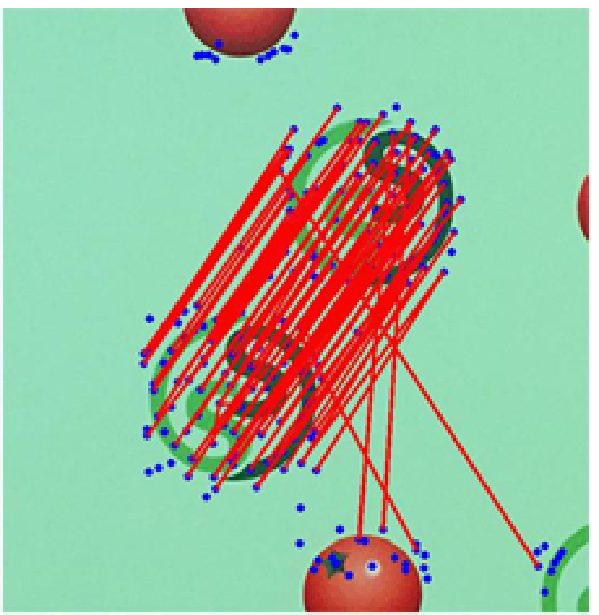}\hspace{0.02cm}
\includegraphics[width=3.5cm,height=3cm]{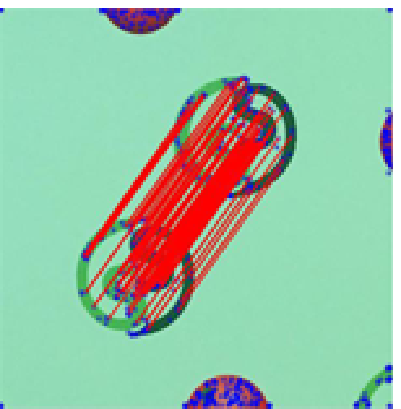}\hspace{0.02cm}

\caption{Top and bottom: exemplar results of SGO and copy-move forgery images. From left to right: (Column 1) Original and tampered images, (Column 2-4) the results based on SIFT\cite{14_amerini2011sift}, SURF\cite{17_bo2010image} and SHFD. The green and blue points in column 2 respectively indicate two clusters, the red lines denote two matching points. }
\vspace{-.1in}
\label{f1}
\end{figure*}

\section{Introduction}
\label{sec:intro}
Copy-move image forgery, as the most commonly occurring forgery type, copies part of the image, and paste it into another part of the same image. Various \emph{Copy-Move Forgery Detection} (CMFD) methods have been proposed, which can be categorized as block-based and key-point-based matching methods.
The first block-based CMFD algorithm by Fridrich \cite{1_fridrich_copymove03}, makes use of Discrete Cosine Transform (DCT) and lexicographical order. Many improved DCT algorithms were subsequently proposed \cite{2_wang2011dwt, 3_6047099}. Muhammad proposed a passive method based on Dyadic Wavelet Transform (DyWT), which combined approximation and detail subbands \cite{4_Muhammad201249}. In addition, some algorithms focus on dimensionality reduction, such as Principal Component Analysis (PCA) \cite{5_popescu2004exposing}. To keep geometric transformation invariance, efforts have been devoted in recent works, such as Fourier-Mellin Transform (FMT) \cite{7_li2010rotation}, invariant moment \cite{8_liu2011passive,10_ryu2010detection}, and Local Binary Patterns (LBP) \cite{12_davarzani2013copy}.
On the category of key-points based methods, Huang proposed CMFD algorithm based on Scale Invariant Feature Transform (SIFT) \cite{13_huang2008detection}, and subsequently, SIFT-improved approaches are proposed\cite{14_amerini2011sift,15_pan2010detecting}. Speeded-Up Robust Feature (SURF) was applied for improving computational efficiency by Xu\cite{17_bo2010image} and Shivakumar \cite{shivakumar2011detection}. Furthermore, some other methods based on DAISY descriptor \cite{18_guo2013duplication} and Harris points\cite{19_chen2013region} are recently proposed.

Existing CMFD methods devote to finding similar
areas to locate tampering, while ignoring that most realistic
scenes are likely to contain \emph{Similar but Genuine Objects} (SGO). With such ambiguity, the performance of CMFD usually degrades when applying to image with SGO. In this work, we proposed a novel CMFD method using \emph{Scaled Harris Feature Descriptors} (SHFD), which performs consistently well and is robust to images containing SGO. Fig.\ref{f1} illustrates an example of CMFD performance degradation using SIFT \cite{14_amerini2011sift} and SURF \cite{17_bo2010image}. Whereas the proposed SHFD method demonstrates promising results. Some important features of our work are as follows,

\begin{enumerate}
\item  Key points are extracted using scaled Harris features, which are scaling invariant. Orientation is assigned to the neighborhood of each key points, in order to achieve rotational invariance.
\item  SHFD performs consistently well for images with \emph{naive}, \emph{rotation}, \emph{scaling} and \emph{free-form distortion} tempering. Empirically, it outperforms SIFT and SURF methods for images from COVERAGE database\cite{24_dataset}. Furthermore, SHFD is robust to post-processing including \emph{blurring}, \emph{noise}, and \emph{jpeg compression} operations.

\end{enumerate}

\section{PROPOSED METHOD}
\label{sec:pagestyle}


There are four main steps in our proposed SHFD algorithm: Scaled Harris points extraction (Section \ref{sec21}), Orientation assignment (Section \ref{sec22}), features extraction (Section \ref{sec23}) and feature matching (Section \ref{sec24}).

\subsection{Scaled Harris points extraction} \label{sec21}

Detecting locations that are invariant
to scale change of the image can be accomplished by
searching for stable features across all scales,
using a continuous function of scale known as scale
space. The pyramid scale space \cite{witkin1984scale} is the most useful model to achieve scale invariance, where octave and interval are used for multi-resolution analysis and image continuity maintenance. Since Harris~\cite{21_harris1988combined} is a classical way to extract corner points with none scale invariance, scaled Harris points are built in our approach by combining pyramid scale.

The pyramid intervals are obtained by Gaussian smoothing and sub-sampling is used to build octaves. For given image $I(x,y)$, the intervals in the same octave are given by: $L(x,y,\sigma)=I(x,y)*G(x,y,\sigma)$, and the first interval of next octave is $L_{oc}(x,y,\sigma) = \text{\textit{sampling}}(L_{oc-1}(x,y,\sigma), \beta)$. Where $G(x,y,\sigma)$ is Gaussian blur, ${\textit{sampling}}$ is the function of down sampling, $\sigma$ and $\beta$ is the scale and sampling factor.

Harris points are classified by using eigen values, $\lambda_1$ and $\lambda_2$ of the second moment matrix $M(x,y)$ as:
\begin{equation}\label{eqn:harris}
M(x,y) = \begin{bmatrix}
  I_x(x,y)^2 & I_x(x,y)I_y(x,y) \\
  I_x(x,y)I_y(x,y) & I_y(x,y)^2 \\
\end{bmatrix}
\end{equation}

where $I_x(x,y)$ and $I_y(x,y)$ respective represents pixel gradient in the $x$ and $y$ direction at point $(x,y)$. If $\lambda_1\approx\lambda_2$ and $\lambda_1\lambda_2\gg0$, the point $(x,y)$ is considered as a corner point. Therefore, corner response can be measured by the following:
\begin{equation}
CR = det(M) - k(tr(M)^2)
\end{equation}
where $det(M)=\lambda_1\lambda_2 = I_x^2I_y^2-(I_xI_y)^2$, $tr(M)=\lambda_1+\lambda_2 = I_x^2+I_y^2$, and $k$ is the weight value, $t\_\{CR\}$ is the threshold value. In our method, the scaled Harris key points are extracted on every scale $L(x,y,\sigma)$ .

\subsection{Orientation assignment} \label{sec22}

To achieve rotation invariance, the most important issue lies in locating the correct neighborhood region. Based on this, the orientation of each key point is a way to help find same region. In our approach, the oriented gradient is used to assign orientation for the neighborhood of each key point. The gradient magnitude and orientation, denoted as $m(x,y)$ and $\theta(x,y)$, are computed on their octaves by using pixel differences as shown in the following equations:

\begin{align}
m(x,y) &= \sqrt{\Delta L_x^2 + \Delta L_y^2 }\\
\theta(x,y) &=\tan^{-1}(\Delta L_y / \Delta L_x)
\end{align}

where $\Delta L_x = L(x+1, y) - L(x-1, y)$, $\Delta L_x = L(x, y+1) - L(x, y-1)$. And then, through dividing $[0, 2\pi]$ to ten regions and getting the histogram of gradient, the maximal magnitude is the orientation of point $(x,y)$. The rotation invariance neighborhood region of point $(x,y)$ is given in the following equation:
\begin{equation}
\begin{bmatrix}
               $\~{x}$ \\
               $\~{y}$ \\
             \end{bmatrix}
             = \begin{bmatrix}
                 \cos\theta & -\sin\theta \\
                 \sin\theta & \cos\theta \\
               \end{bmatrix} \begin{bmatrix}
               x \\
               y \\
             \end{bmatrix}
\end{equation}
where $\begin{bmatrix}
               x \\
               y \\
             \end{bmatrix}$ represent pixels in a 4$\times$4 square neighborhood among the center $(x,y)$,
             $\begin{bmatrix}
               $\~{x}$ \\
               $\~{y}$\\
             \end{bmatrix}$ are the coordinates of oriented region, and $\theta$ is the orientation of key point $(x,y)$.

\subsection{Feature descriptor extraction} \label{sec23}

Accurate feature descriptions are able to capture the weeny image details are the key of robustness on SGO. Local Binary Patterns (LBP) is an effective texture feature descriptor due to its low computational complexity, invariance to monotonic gray-scale changes and texture description ability\cite{22_ojala2002multiresolution}. Besides, DCT is a well known sparsifying transform for image regions where information is highly concentrated in its low-frequency component. Additionally, Singular Value Decomposition (SVD) is the tool commonly used in the dimensionality reduction methods. Therefore, coefficients in the DCT domain, and singular values of the image data are good features which are robust to noise and interference.

In our method, Suppose $M$ is the neighborhood region of point $(x,y)$, which is a 4$\times$4 square matrix. Uniform LBP $(LBP_{P,R}^{u2})$ and the rotation invariant uniform LBP $(LBP_{P,R}^{riu2})$ keep the rotation invariance through starting from minimum LBP value and remove the redundancy, where $P$ is the number of pixels in neighborhood on a circle of radius $R$.  DCT coefficient could be extracted from $M$ and reshaped as a vector $dct$ with dimensions of 16. The diagonal entries of singular matrix $M$ in descending order are recorded as SVD vectors $svd$.
In total, the descriptor of point $(x,y)$ is presented with four feature vectors $V=[V_{1}(LBP_{8,1}^{u2})$, $V_{2}(LBP_{16,2}^{riu2})$, $V_{3}(dct)$, $V_{4}(svd)]$ with dimensions of 93($V_1(59)+V_2(14)+V_3(16)+V_4(4)$).


\subsection{Feature matching} \label{sec24}
Since the next octave is the down sampling by factor $\beta$, the equation of mapping the Harris point $(x, y)$ to original image $(X, Y)$ is:
\begin{equation}
X = x\times(1/\beta)^{oc-1},  \hskip 0.3cm Y = y\times(1/\beta)^{oc-1}
\end{equation}

If Euclidean distances of every mixed feature are less than the threshold $\epsilon$, the point pairs $(i,j$) are regarded as candidate matching pairs. After this we obtain the matrix of matching pairs, recorded as follows:
\begin{equation}
match(i,j) = [(X_i, Y_i), (X_j, Y_j)]
\end{equation}

Since there is a significant amount of false matches, a way of removing false matches is applied. RANdom SAmple Consensus (RANSAC)\cite{23_RANSAC} is an iterative method to remove false matches. In our method, RANSAC evaluates a translation matrix model on the dataset $match(i,j)$ and removes false matches that are not compatible with it.

\begin{figure} [b]
\centering
\includegraphics[width=4cm,height=2.8cm]{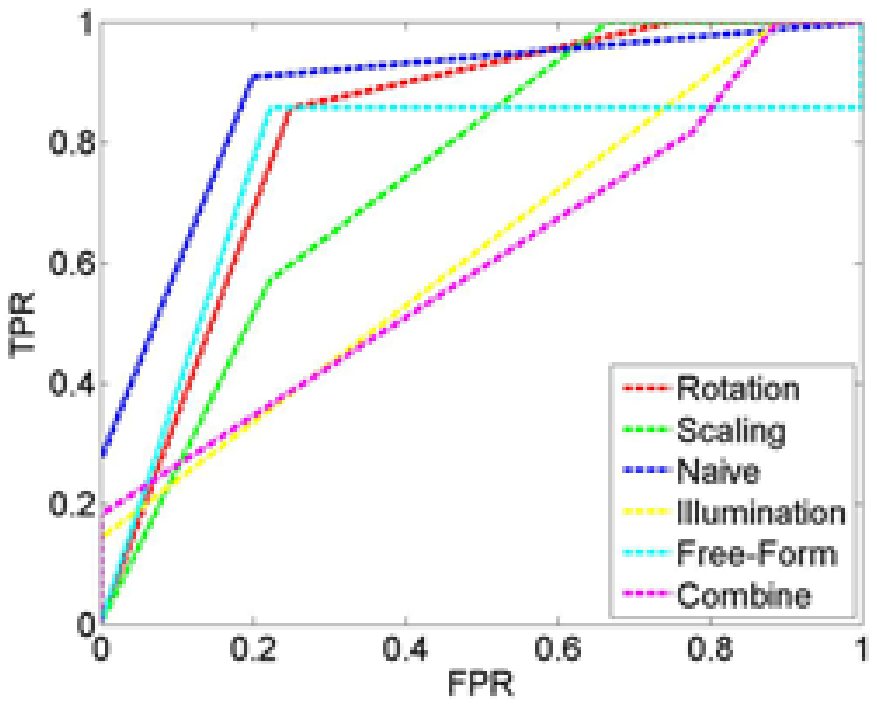}\hspace{0.01cm}
\includegraphics[width=4cm,height=2.8cm]{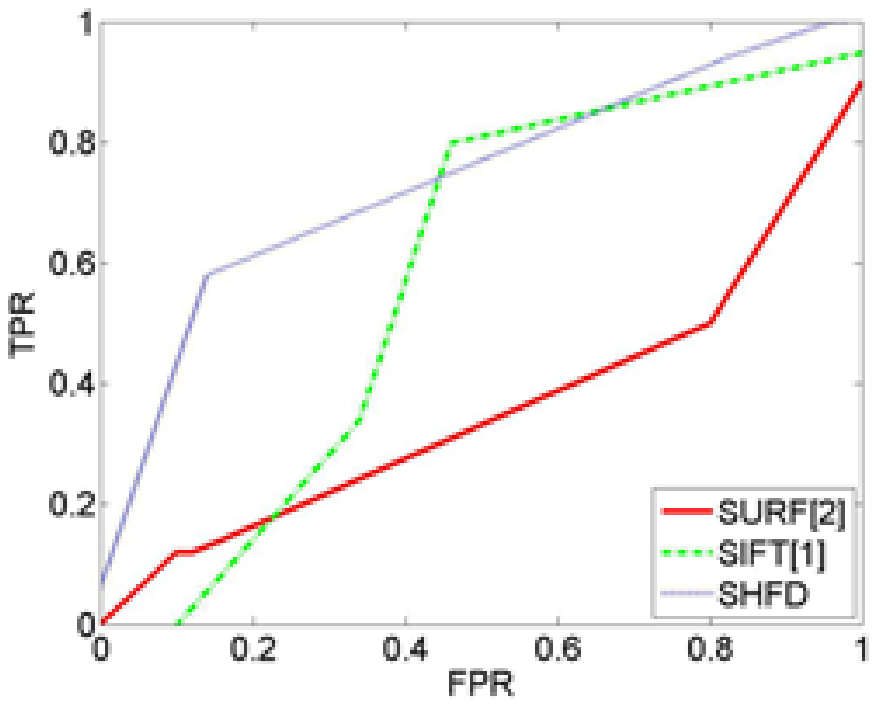}
\caption{From left to right: ROC curves respect to tampering factors and COVERAGE database.}
\vspace{-.2in}
\label{f2}
\end{figure}

\begin{figure*} [htb]
\centering
\includegraphics[width=3.5cm,height=2.8cm]{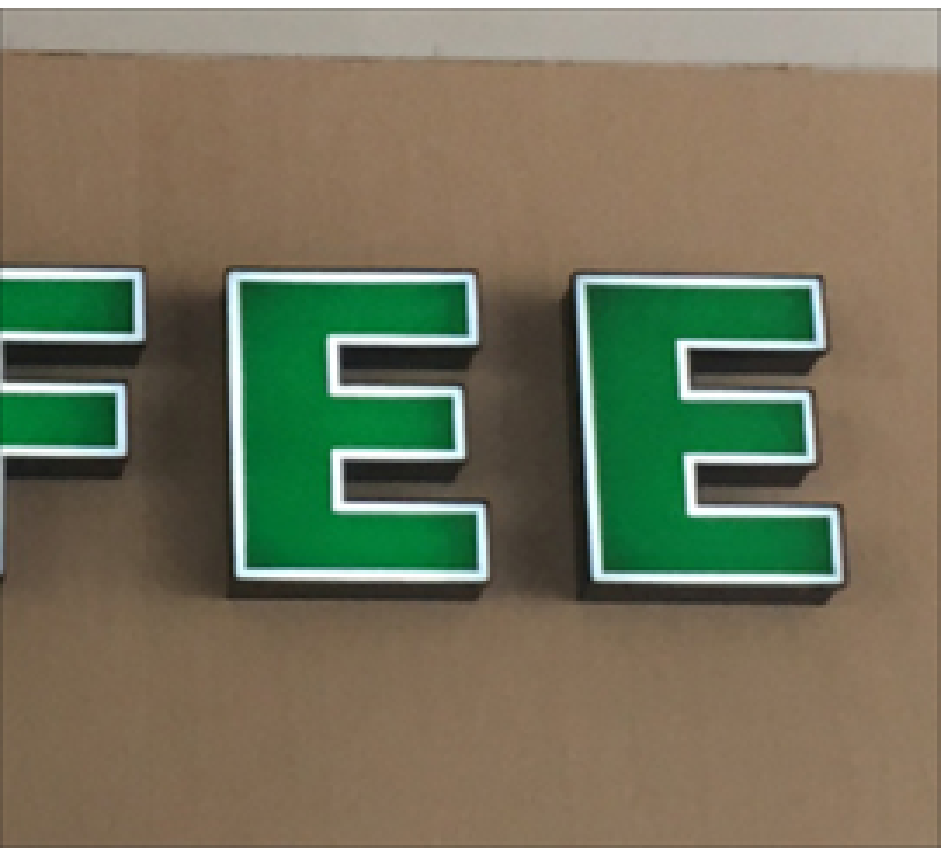}\hspace{0.01cm}
\includegraphics[width=3.5cm,height=2.8cm]{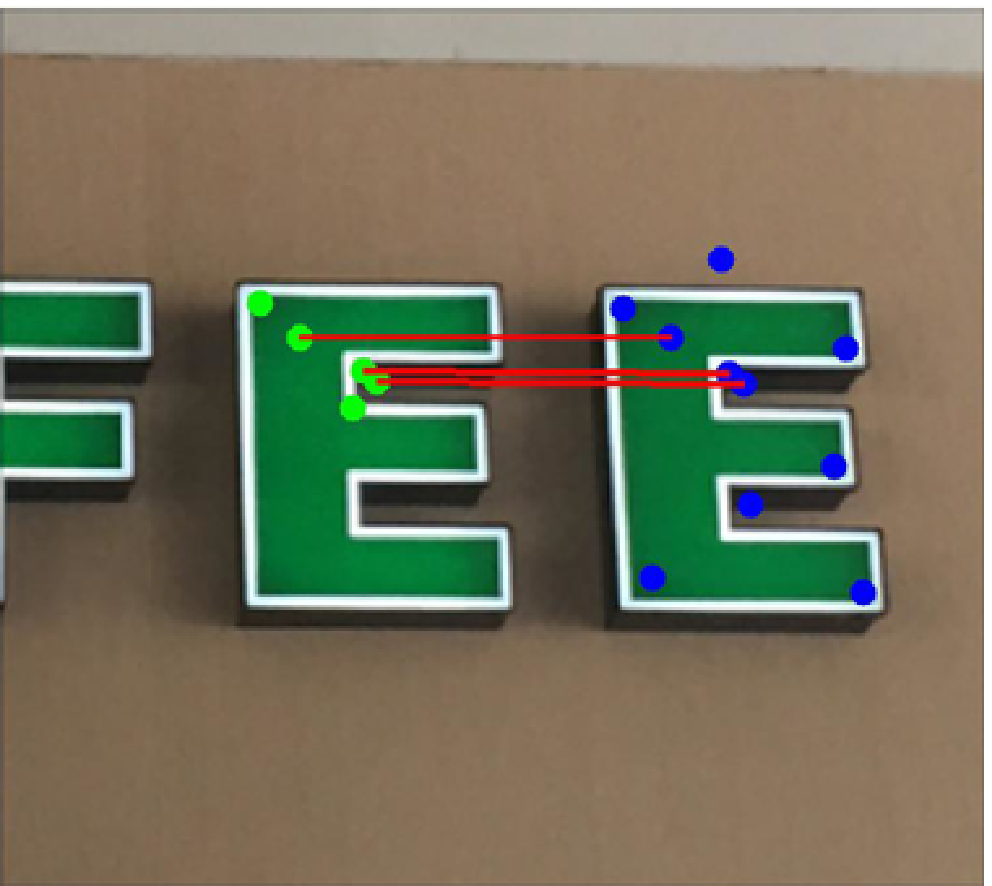}\hspace{0.01cm}
\includegraphics[width=3.5cm,height=2.8cm]{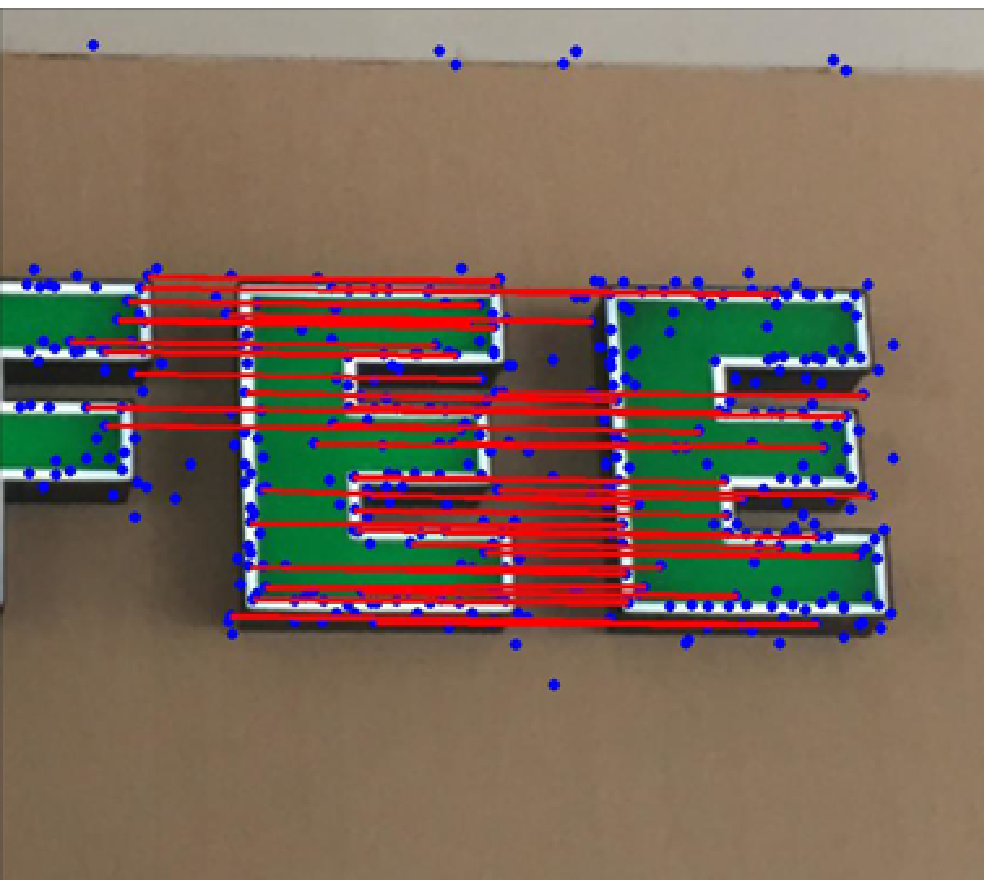}\hspace{0.01cm}
\includegraphics[width=3.5cm,height=2.8cm]{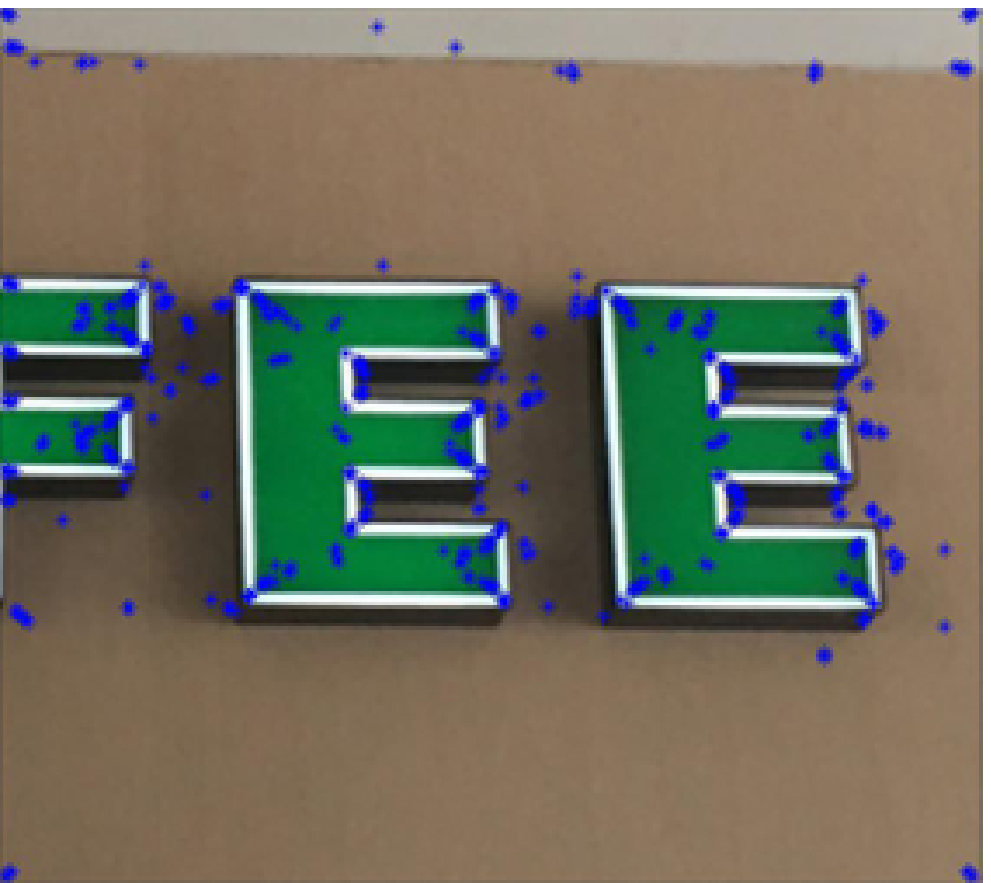}\hspace{0.01cm}
\vspace{.1cm}

\includegraphics[width=3.5cm,height=2.8cm]{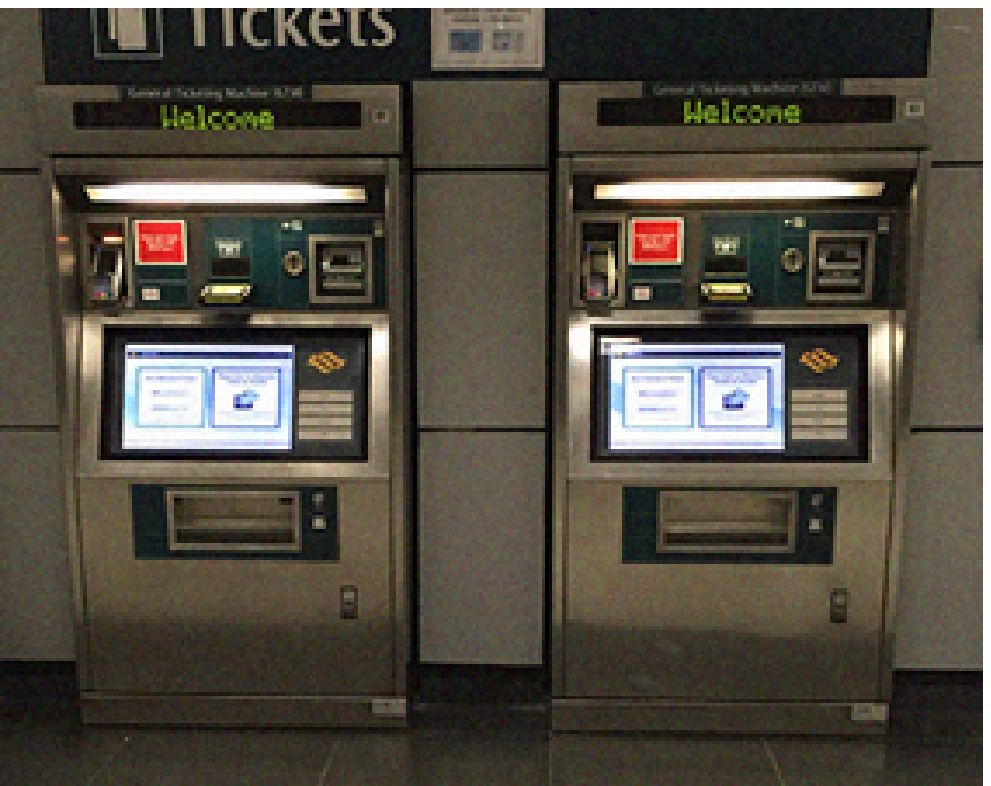}\hspace{0.01cm}
\includegraphics[width=3.5cm,height=2.8cm]{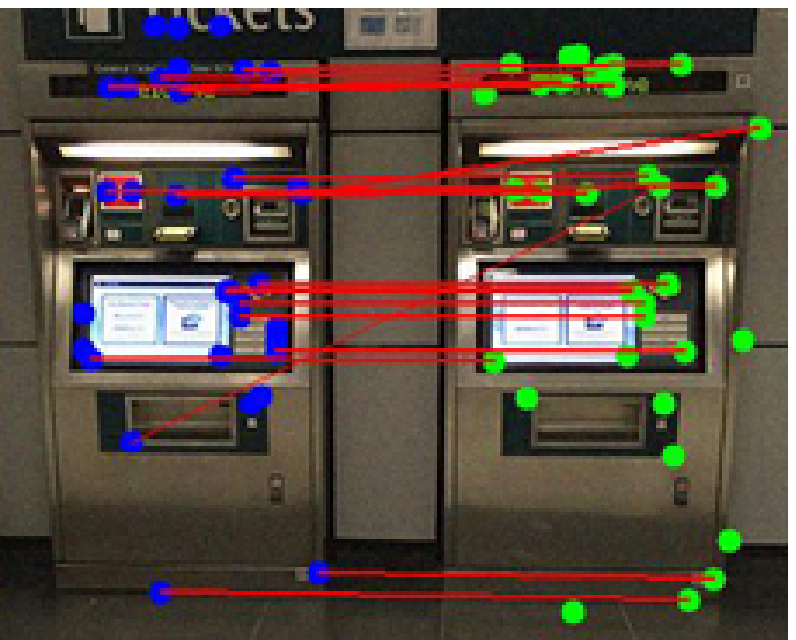}\hspace{0.01cm}
\includegraphics[width=3.5cm,height=2.8cm]{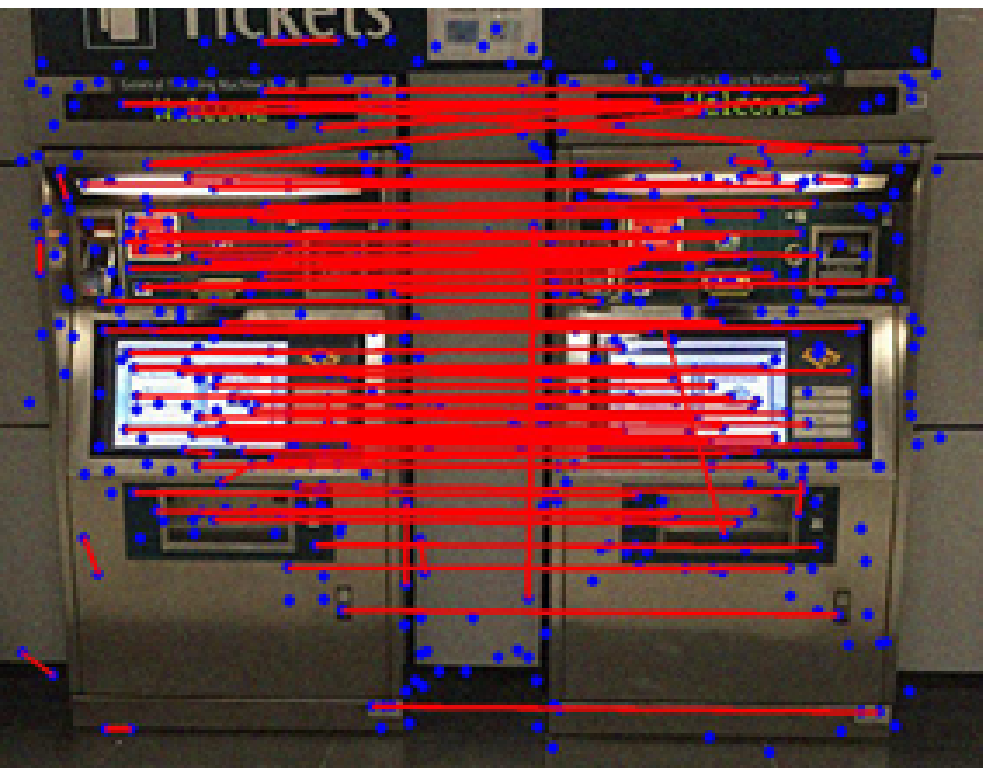}\hspace{0.01cm}
\includegraphics[width=3.5cm,height=2.8cm]{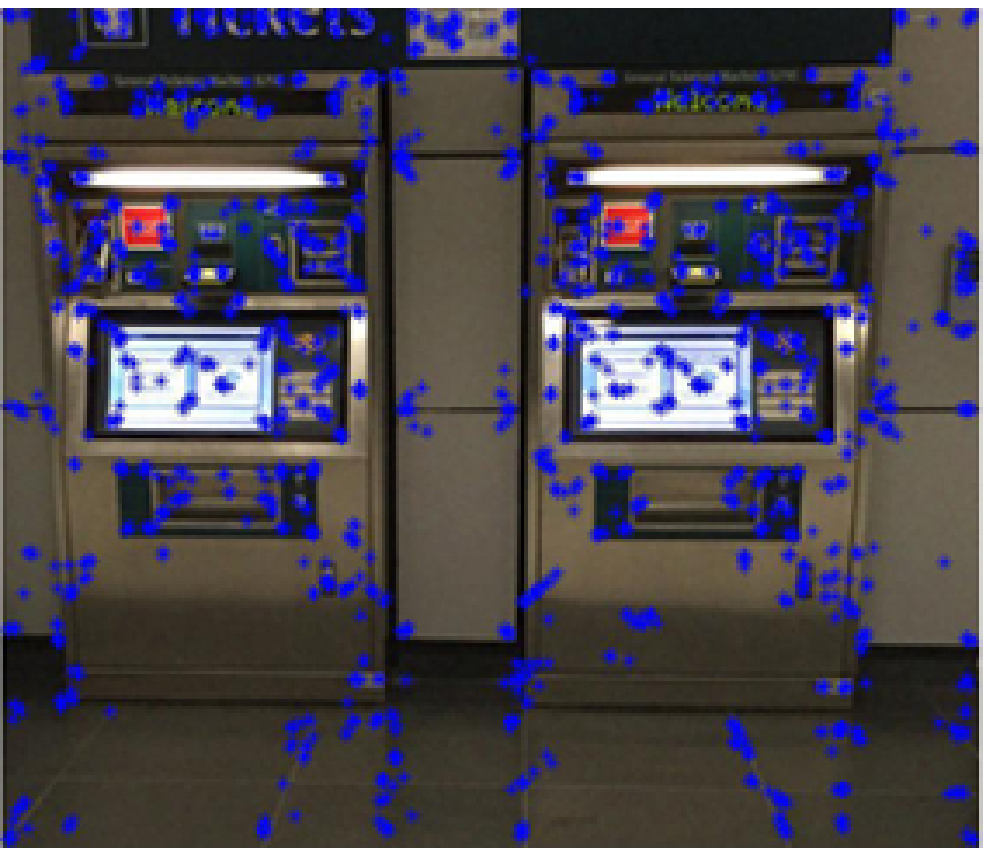}\hspace{0.01cm}
\vspace{.1cm}

\includegraphics[width=3.5cm,height=2.8cm]{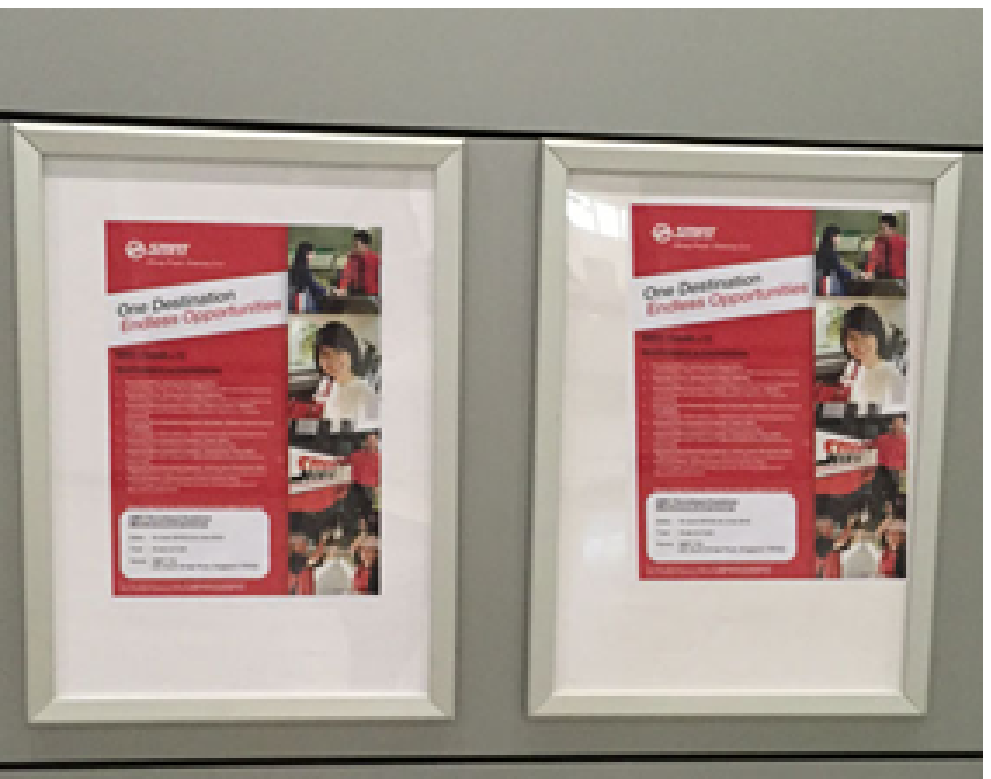}\hspace{0.01cm}
\includegraphics[width=3.5cm,height=2.8cm]{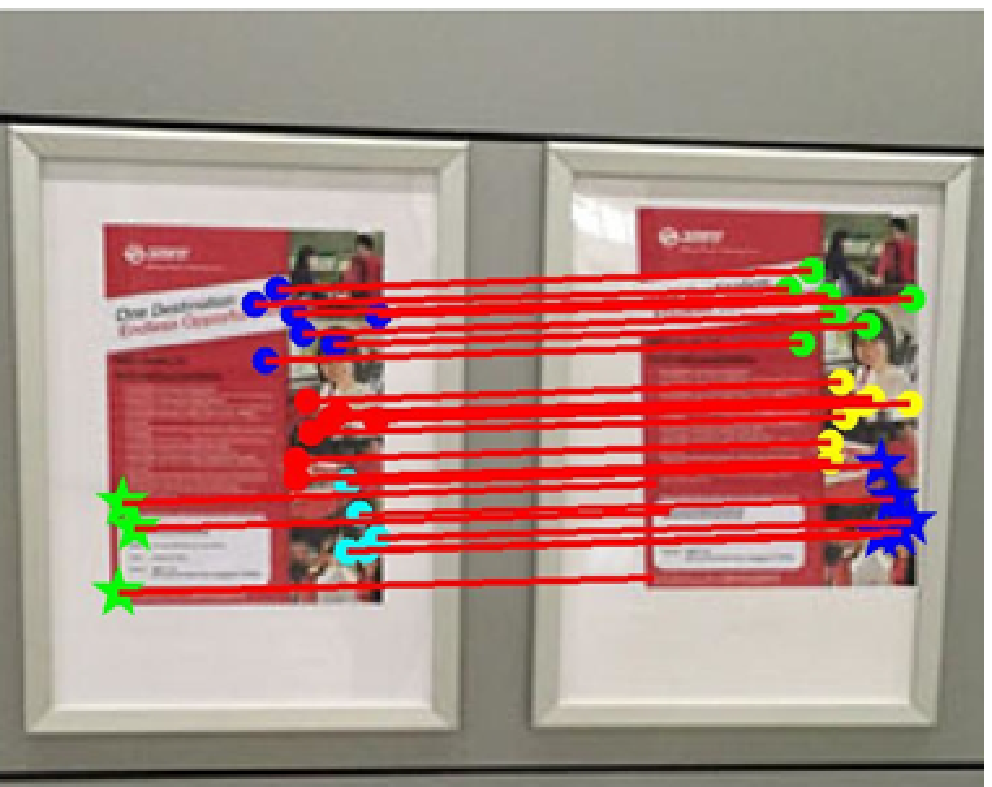}\hspace{0.01cm}
\includegraphics[width=3.5cm,height=2.8cm]{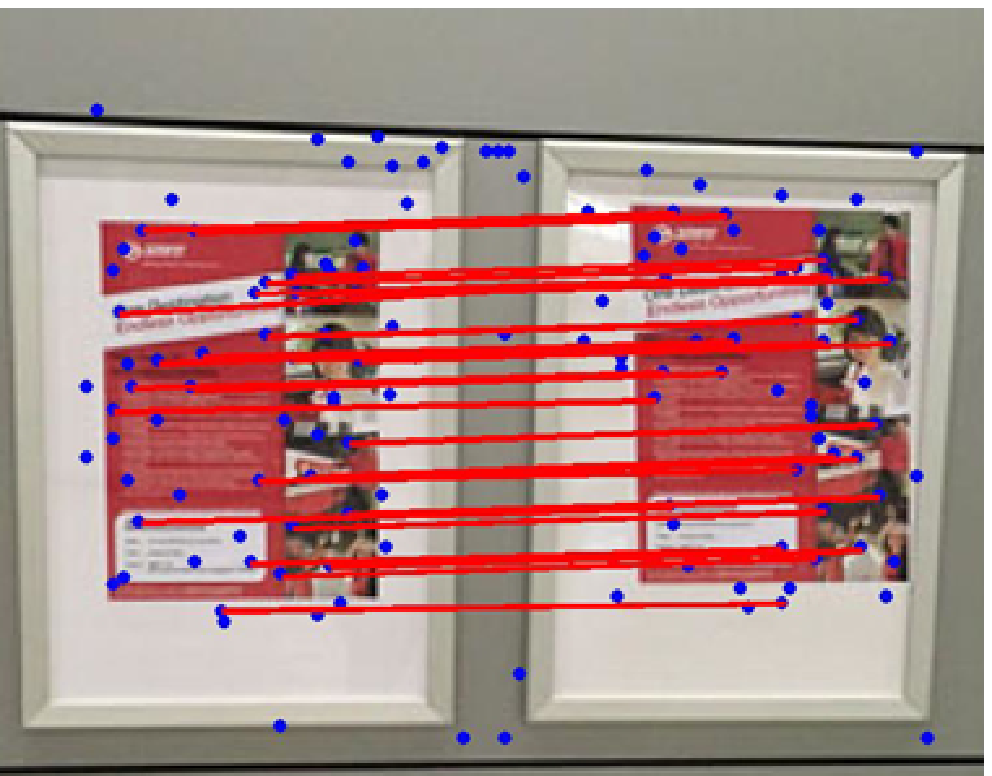}\hspace{0.01cm}
\includegraphics[width=3.5cm,height=2.8cm]{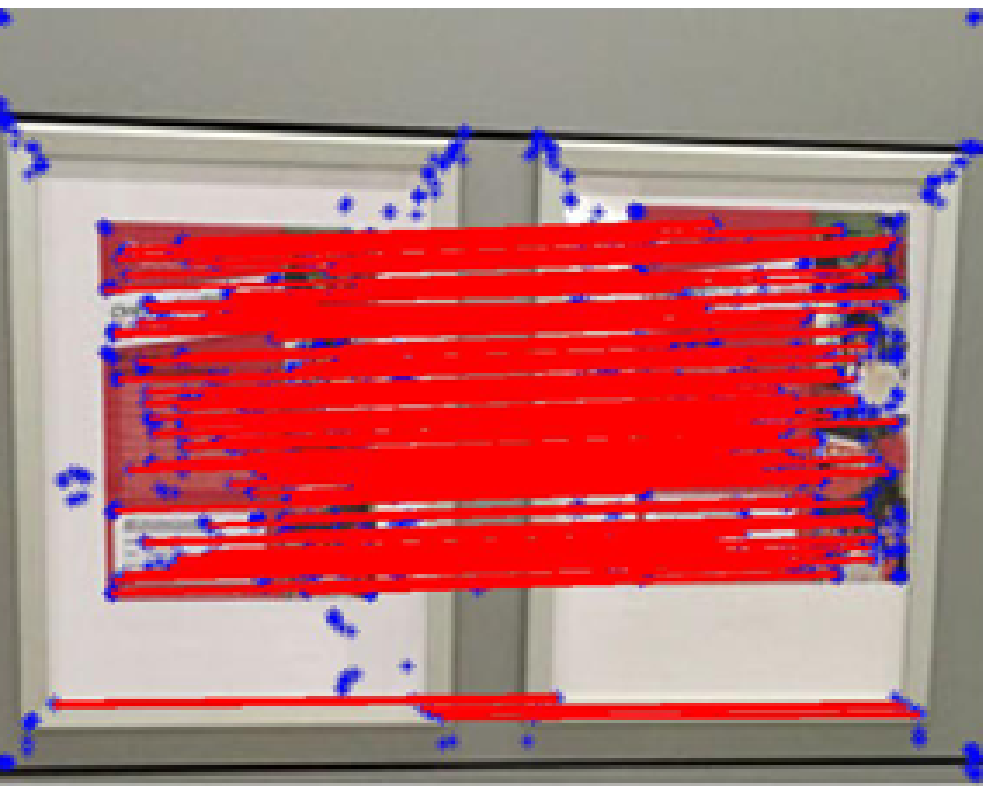}\hspace{0.01cm}
\vspace{.1cm}
\caption{From top to bottom: Exemplar results of two SGO and one tampered images varying with respect to \emph{blurring}, \emph{noise} and \emph{jpeg compression}. From left to right: (Column 1) tested images, (Column 2-4) the results based on SIFT\cite{14_amerini2011sift}, SURF\cite{17_bo2010image} and SHFD.}
\vspace{-.1in}
\label{f3}
\end{figure*}

\begin{figure*} [htb]
\centering
\includegraphics[width=5.5cm,height=3.5cm]{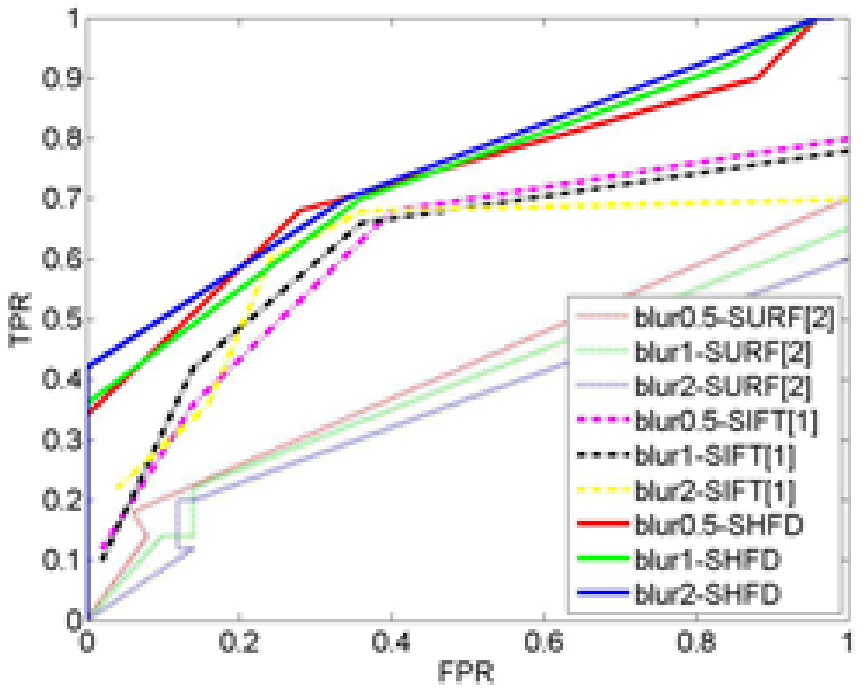}\hspace{0.01cm}
\includegraphics[width=5.5cm,height=3.5cm]{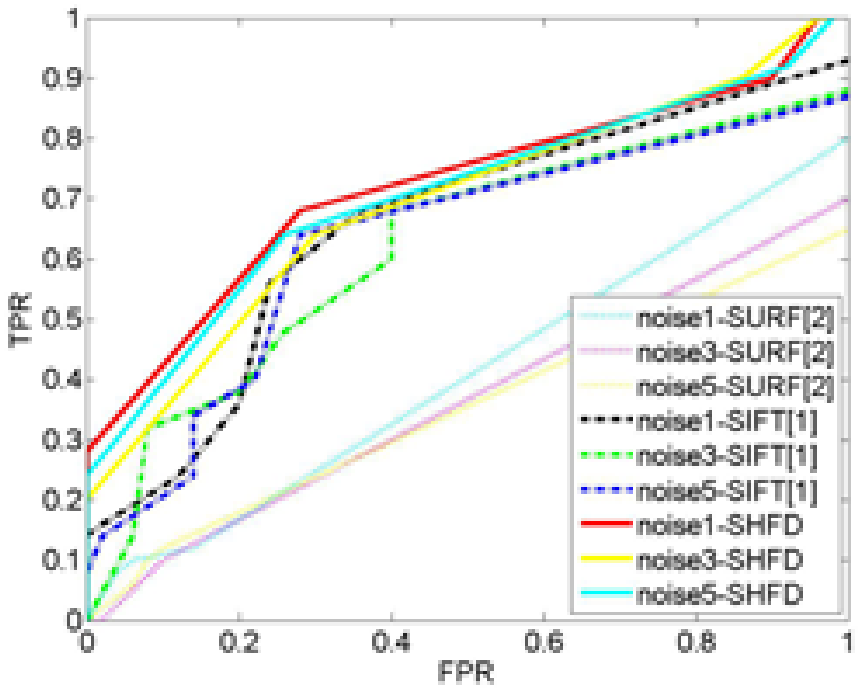}\hspace{0.01cm}
\includegraphics[width=5.5cm,height=3.5cm]{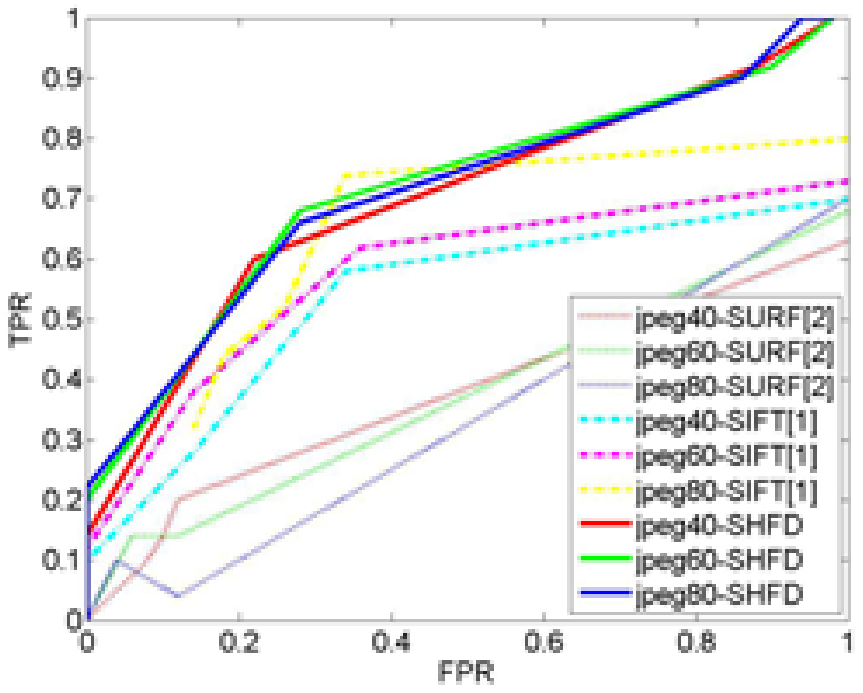}\hspace{0.01cm}
\vspace{.1cm}
\setlength{\abovecaptionskip}{-1.3cm}
\caption{From left to right: ROC curves varying with respect to \emph{blurring}, \emph{noise} and \emph{jpeg compression}}
\vspace{-.1in}
\label{f4}
\end{figure*}

\section{EXPERIMENTS}

\subsection{Database}
We evaluate the proposed SHFD algorithm, as well as popular SIFT\cite{14_amerini2011sift} and SURF \cite{17_bo2010image} methods over the newly proposed COVERAGE database \cite{24_dataset}. The original images of the forgery pairs from COVERAGE all contain SGO. The selected images are forged with 6 different tampering factors, respectively \emph{naive}, \emph{rotation}, \emph{scaling}, \emph{illumination}, \emph{free-form distortion} and \emph{combined factors}.

\subsection{Parameters and Metrics}

The input parameters required by the
methods are set as follows: $oc=4$ (number of pyramid octaves), $in=4$ (number of pyramid intervals), $\beta=1.25$ (sampling factor in pyramid space), $t\_\{CR\} =0.02*max(CR)$ (threshold for the corner response) and $k=0.05$ (weight in corner response).

To evaluate the performance of CMFD on images with SGO, true positive rate (TPR) and false positive rate (FPR) are used as evaluation metrics. They are defined as follows,
\begin{align}
TPR &= \frac{\#\text{image detected as forgery being forgery}}{\#\text{forgery images}}\\
FPR &= \frac{\#\text{image detected as forgery being origin}}{\#\text{origin images}}
\end{align}

\subsection{Numerical Results}

In this part, we compare empirical performance using the proposed SHFD algorithm to popular CMFD methods, including SIFT \cite{14_amerini2011sift}, and SURF \cite{17_bo2010image} algorithms. We present and analyze the numerical results with different tampering factors and post-processing methods.

\subsubsection{Performance on tampering factors discussion}

To minimize the visible traces of forgery, various types of tempering factors are applied in the forged image. We now analyze how CMFD performances varied with different tempering factors. We evaluate TPR and FPR values by applying SHFD algorithm to 100 images from COVERAGE. Fig. \ref{f2} in the left side illustrates ROC curves subject to different tampering factors. Empirically, our proposed SHFD algorithm performs consistently well for images with \emph{naive}, \emph{rotation}, \emph{scaling} and \emph{free-form distortion} tempering. However, we also observed reasonable performance degradation for images with complicated tempering factors such as \emph{illumination}, and \emph{combined factors}. More sophisticated features which impose illumination variance are required to handle more complex tempering factor in the future work.

To compare the proposed SHFD method to the popular SIFT and SURF methods, we also evaluate TPR and FPR values using SIFT\cite{14_amerini2011sift}, and SURF \cite{17_bo2010image} methods over COVERAGE database. The overall ROC curves are plotted in the right of Fig.\ref{f2}. The empirical results obtained by the proposed SHFD method demonstrate promising performance compared to SIFT and SURF methods.

\subsubsection{Post-processing experiments}

It is important to study the CMFD behavior subject to post-processing operations such as blurring, noise corruption and JPEG compression, since similar effects usually occur during image transmission and processing. We now artificially edit images from COVERAGE databases with operations including Gaussian blurring (window size, $w=3$, and sigma, $\sigma=[0.5, 1,2 ]$), Gaussian noise corruption (mean, $m=0$, and variance, $var=[1, 3, 5]$) and JPEG compression with a decreasing quality factor of $[80, 60, 40]$. Fig.\ref{f3} shows three examples of processed images with blurring, noise corruption and JPEG compression, as well as their CMFD matching results using SIFT, SURF and SHFD methods.
The corresponding ROC curves are plotted in Fig.\ref{f4} with blurring, noise corruption and JPEG compression respectively. From the plotted curves, we observe promising robustness of the proposed SHFD method subject to \emph{blurring}, \emph{noise}, and \emph{jpeg compression} operations.

\section{CONCLUSIONS}

Detecting images with SGO and copy-move is a common issue in CMFD. However, it is often overlooked in existing methods. In this paper, an efficient method called SHFD was proposed and evaluated against two state-of-art methods. The results show that SHFD is the only one that could distinguish images with SGO and copy-move forgery. Furthermore, it also determines the geometric transformations and post-processing applied to the forged regions. However, our method preforms unsatisfied in the illumination variance, which will be continued to work in the future study.

\label{sec:refs}

\bibliographystyle{IEEEbib}
\bibliography{icassp_v2_Bihan}

\end{document}